\documentclass[11pt,a4paper]{article}
\pdfoutput=1
\usepackage{jheppub}
\usepackage{amsmath}
\usepackage{graphicx,tabularx}
\usepackage{url}
\usepackage{times}
\usepackage{ulem}
\usepackage{color}

\def\tablename{Table}
\def\figurename{Figure}

\def\nj{n_\text{jets}}

\newcommand\one{\leavevmode\hbox{\small1\normalsize\kern-.33em1}}

\newcommand{\ord}{\mathcal{O}}

\newcommand{\qqqquad}{\qquad \qquad \qquad}

\newcommand{\gev}{{\ensuremath\rm GeV}}

\def\slashchar#1{\setbox0=\hbox{$#1$}           
   \dimen0=\wd0                                 
   \setbox1=\hbox{/} \dimen1=\wd1               
   \ifdim\dimen0>\dimen1                        
      \rlap{\hbox to \dimen0{\hfil/\hfil}}      
      #1                                        
   \else                                        
      \rlap{\hbox to \dimen1{\hfil$#1$\hfil}}   
      /                                         
   \fi}

\def\eg{{\sl e.g.} \,}
\def\ie{{\sl i.e.} \,}

\newcommand{\be}{\begin{eqnarray*}}
\newcommand{\ee}{\end{eqnarray*}}
\newcommand{\gl}[1]{(\ref{#1})}

\newcommand{\bee}{\begin{eqnarray}}
\newcommand{\eee}{\end{eqnarray}}
\newcommand{\beeq}{\begin{equation}}
\newcommand{\eeeq}{\end{equation}}

\begin{document}

\title{Establishing Jet Scaling Patterns with a Photon}

\author[a]{Christoph Englert} 

\author[a]{, Tilman Plehn} 

\author[a]{, Peter Schichtel} 

\author[a,b]{, and Steffen Schumann} 

\affiliation[a]{Institut f\"ur Theoretische Physik, Universit\"at
  Heidelberg, Germany} 
\affiliation[b]{II. Physikalisches Institut,
  Universit\"at G\"ottingen, 37077 G\"ottingen, Germany}

\abstract{Staircase and Poisson scaling are two typical patterns we
  observe for the exclusive number of jets at high energy hadron
  colliders. We examine these scaling properties for photon plus jets
  production at the LHC and find that this channel is well suited to
  study these features. We illustrate and discuss when to expect each
  of the two patterns, how to induce a transition through kinematic
  cuts, and how photons are related to heavy gauge
  bosons. Measurements of photon+jets production is therefore
  providing valuable information on exclusive jet scaling, which is
  going to help to eventually understand the theoretical origin of
  exclusive jet scaling properties in more detail.}
\maketitle

\section{Introduction}

After testing and reproducing many interesting aspects of the Standard
Model at the LHC, the focus of the ATLAS and CMS collaborations is
rapidly moving toward searches for Higgs particles~\cite{higgs} or
physics beyond the Standard Model~\cite{review}. The production rates
for any of these search channels are small, for example compared to
$W/Z$+jets or top pair production, channels which constitute their
main backgrounds. Traditionally, at hadron colliders we have relied on
the appearance of leptons, photons or missing transverse energy to
point us to interesting new physics processes. In this approach QCD
effects and jet production are either ignored or considered a
nuisance.

Starting with the suggested searches for a light Higgs boson in weak
boson fusion~\cite{wbf}, this attitude has changed; this search shows
how the QCD structure of signal events can be turned into a powerful
handle to reject large backgrounds. The key analysis tools are
(central) jet vetos~\hbox{\cite{cjv,prl,andy}}, which for example
suppress QCD-initiated $W/Z$+jets events or hadronic top pair
production. Implicitly, this approach is adopted in Higgs searches for
example in the $H \to WW$ or $H \to \gamma \gamma$ channels, when
those searches are divided into 0-jet, 1-jet and soon 2-jet
strategies~\cite{higgs}.

In a similar spirit, searches for example for supersymmetry benefit
from the measurement of the number of jets which includes information
on the color structure of the new heavy states~\cite{autofocus}; the
only caveat is that we need to carefully separate decay jets from QCD
jet radiation associated with hard processes~\cite{autofocus,jet_rad}.
What is missing for all such analyses is an experimentally established
and theoretically sound link between choosing $n$-jet samples for an
analysis and a systematic study of the corresponding $\nj$
distribution for signal and background processes~\cite{prl}. A
dedicated study along this line would map out the behavior of
exclusive $\nj$ distributions after different cuts, understand its
basic features, and quantify the notorious theory uncertainties
associated with jet counting.  As it will turn out, the production of
a photon in association with QCD jets is a perfect basis for such
studies and complements the ones, which have already been performed at
the LHC \cite{atlasjets}.  \bigskip

From text book quantum field theory we know that successive
$p_T$-ordered photon radiation off a hard electron --- as well as
successive gluon radiation off a hard quark --- follows a Poisson
pattern for the {\sl exclusive number} of photons or
gluons~\cite{peskin}. This pattern corresponds to a simple
probabilistic picture of successive independent splitting. The
splitting probability is linked to the coupling constant, the color
factor, the form of the splitting kernel, and a scale logarithm.  The
Poisson scaling pattern, however, has not been observed in inclusive
production at hadron colliders since UA1/UA2. Instead, for many
processes we find staircase scaling, namely a constant ratio of
exclusive $n$-jet rates
$\sigma_{n+1}/\sigma_n=R$~\cite{staircase1,staircase2}. This feature
has been studied by the LHC experiments \cite{recentscaling} and also
finds application in background-modelling in phenomenological
approaches \cite{backgroundsc}.  \bigskip

The description of exclusive jet rates is at odds with our description
of QCD at hadron colliders. Parton densities obeying the DGLAP
equation \cite{dglap} resum collinear logarithms and absorb the
corresponding infrared divergences. As a consequence, any computation
based on such parton densities is jet inclusive, \ie it allows for an
unspecified number of collinear jets radiated off the incoming
partons. Strictly following the DGLAP approximation, these perfectly
collinear jets are not observable at the LHC. However, the assumption
of perfect collinearity is modified by the initial state parton
shower, which redistributes the dominantly collinear jet radiation
into the physical phase space. Evaluating exclusive event samples with
exactly $n$ jets is theoretically limited by the precision of the
parton shower description, including its obvious breakdown for
sizeable transverse momenta~\cite{scet}. This is the reason why in the
past exclusive $\nj$ distributions could rarely be exploited to
compare collider data to QCD predictions. However, as described above,
current LHC analyses force us to overcome this limitation and study
exclusive $\nj$ distributions (including cuts corresponding to jet
veto survival probabilities) starting from perturbative QCD.

Matching of a hard QCD matrix element with a collinear parton
shower~\cite{early_matching,ckkw,mlm,lecture} allows us to reliably
simulate and study jet scaling patterns from perturbative QCD. It does
not require a fundamental re-organization of QCD perturbation
theory~\cite{simone} but simply relies on the proper phase space
simulation of collinear logarithms.  This way it does not only include
the radiation of one or two hard jets, as correctly described by
fixed-order QCD calculations, but any number of radiated jets
including high multiplicities obviously well described by the parton
shower.  We use {\sc Sherpa}~\cite{sherpa} with its {\sc
  Ckkw}~\cite{ckkw} matching scheme to describe radiation of up to
seven jets with high precision. Typically, we check an additional two
more jets for unexpected features, but with correspondingly reduced
statistics.\bigskip

As we will show in this paper, a particularly promising channel to
measure exclusive jet rates and compare them to QCD predictions is
hard photon production in association with jets.

The cross section is large enough to already have enough data to not
only validate Monte Carlos, but also test potential $\nj$ scaling
hypotheses in various phase space regions. We will show how to define
and extract different kinematical regimes of the photon to compare
various hypotheses of QCD radiation with data and to cross check jet
emission against other production modes. Possible applications towards
new physics searches under the premise that Poisson or staircase
scaling are phenomenologically realized to good approximation have
been discussed in Refs.~\cite{higgs,cjv,prl}.

\section{Staircase scaling}
\label{sec:staircase}

Some qualitative implications of staircase scaling are already known
from the analysis of the inclusive $\nj$ distribution of Standard
Model processes.  These include pure QCD jets
production~\cite{autofocus,ex_lhc} and $W/Z$ production with
jets~\cite{staircase1,staircase2,ex_lhc}.  As we will see,
$\gamma$+jets production when constrained in certain cut scenarios
provides an additional channel, hence contributing to a better
understanding of the possibly observed $\nj$ scaling patterns.
Staircase scaling is defined in terms of constant ratios of the
experimentally measured jet-inclusive $n$-jet cross sections
\begin{equation}
  \hat{R}_{(n+1)/n}=\frac{\hat{\sigma}_{n+1}}{\hat{\sigma}_n} \equiv \hat{R} \,.
  \qqqquad \text{(jet-inclusive)}
\end{equation}
It turns out that we can equivalently formulate this condition in
terms of inclusive ($\hat{\sigma}$) and exclusive ($\sigma$) numbers
of jets. Correspondingly defining
\begin{equation}
  R_{(n+1)/n} = \frac{\sigma_{n+1}}{\sigma_n} \equiv R
  \qqqquad \text{(jet-exclusive)}
\label{eq:staircase_def}
\end{equation}
for the exclusive rates, the resulting ratios are
identical~\cite{autofocus},
\begin{alignat}{5}
  \hat{R}
  &= \dfrac{\sum_{j=n+1} \sigma_j}{\sigma_n + \sum_{j=n+1} \sigma_j} =
  \dfrac{\sigma_{n+1} \sum_{j=0}^\infty R^j}{\sigma_n + \sigma_{n+1}
    \sum_{j=0}^\infty R^ j} = \dfrac{R \sigma_n \,
    \dfrac{1}{1-R}}{\sigma_n + R \sigma_n \, \dfrac{1}{1-R}}
  &= R \; .
  \label{eq:staircase_excl}
\end{alignat}
This way the merits of perturbative QCD and its perturbative
predictions directly translate to the jet-exclusive final states in a
well-defined approach where lower multiplicities are utilized to
constrain the higher ones.

While a proper analytical derivation of this feature from first
principles is still missing, it is observed to a good approximation in
both experimental data and theoretical calculations using matrix
element and parton shower merging \cite{ex_tev,ex_lhc}. Only using jet
radiation via parton showering this scaling feature is not matched as
well, which is expected given that such inclusive processes do not
offer a hard scale in relation to which we can define collinear
radiation. In addition, for $W$+jets production it has been shown that
fixed order QCD corrections stabilize the observed staircase
pattern~\cite{blackhat}.

\bigskip

Unlike the other cases mentioned above, at first sight photon
production does not posses an obvious jet scaling behavior. It only
occurs once we include strong separation cuts between the photon and
each of the jets, effectively removing any logarithmic enhancement
linked to QED photon radiation.  Moreover, as we will show in this
paper, different basic cuts can easily induce different scaling
patterns.

For our simulation we rely on {\sc Sherpa} v1.3.0~\cite{sherpa} and
its {\sc Ckkw} matching up to five matrix element jets. We reconstruct
jets using the anti-$k_T$ algorithm from {\sc
  FastJet}~\cite{fastjet,antikt} with $R_{\text{anti}-k_T}=0.4$, which
gives us a very moderate geometric separation of two jets. When
dealing with photons in a QCD environment some familiar subtleties
have to be considered~\cite{frixione,photon}: a photon can arise from
non-perturbative fragmentation. Those photons are not useful in our
case since our focus is obviously not on QED corrections to multi-jet
production rates.  Therefore, we opt for a solid photon isolation. A
naive hard cut \eg on the jet-photon $R$ distance limits the phase
space of soft gluon emission and is infrared unsafe. We instead define
an isolated photon through a hadronic energy deposit of less than 10\%
of $p_{T,\gamma}$ in a cone of size $R<0.4$ centered around the photon
direction~\cite{frixione}.  If this criterion is not met the photon
candidate is pushed into the jet finding algorithm.\bigskip

For reconstructed jets and photons in this section we then require
\begin{equation}
  p_{T,\gamma} > 50~\gev, 
  \quad 
  p_{T,j} > 50~\gev, 
  \qqqquad 
  |\eta_\gamma| < 2.5, 
  \quad 
  |y_j| < 4.5\,,
\label{eq:staircuts}
\end{equation}
where $\eta$ and $y$ denotes the pseudo-rapidity and rapidity,
respectively. These cuts are very inclusive and democratic, so we can
expect to observe the staircase scaling behavior known for pure QCD
jets.

\begin{figure}[t]
  \includegraphics[width=0.46\textwidth]{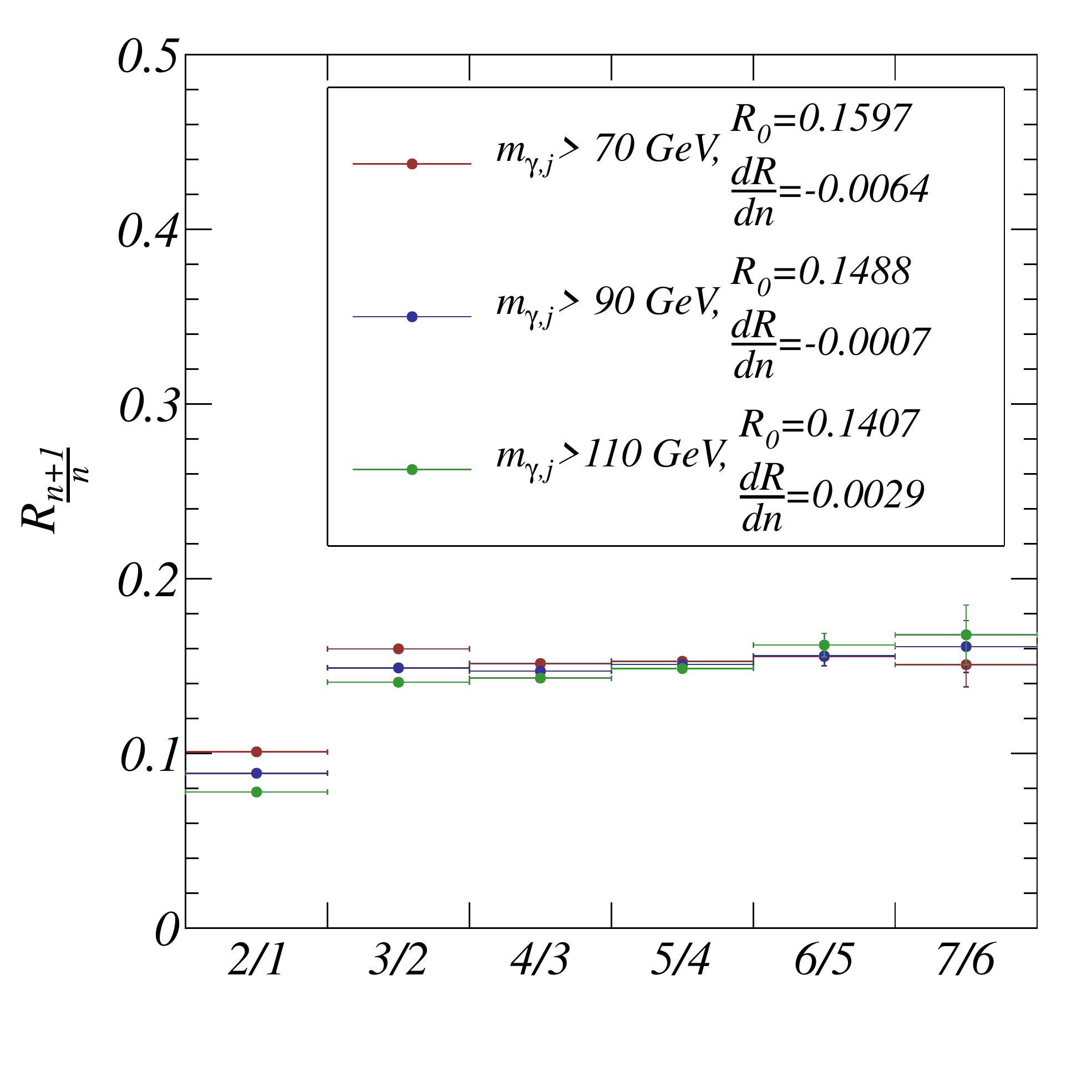}
  \hspace*{0.05\textwidth}
  \includegraphics[width=0.46\textwidth]{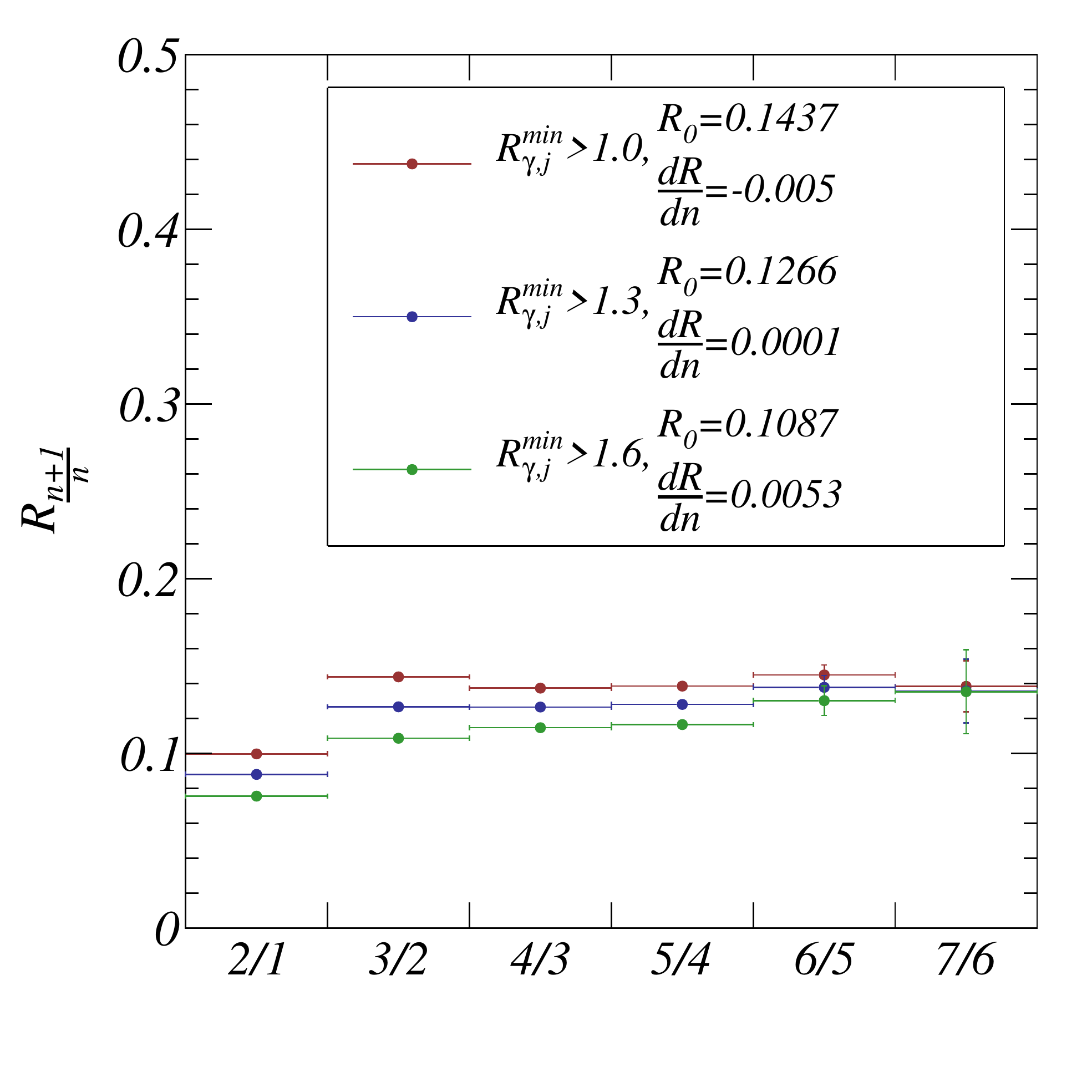}
  \caption{Two scenarios establishing staircase scaling for
    $\gamma$+jets production at the 7~TeV LHC. Left: invariant mass
    criterion of $m_{\gamma j}>70,~90, 110$~GeV for each jet. Right:
    geometric separation between the photon and each of the jets.  The
    extracted values for $R_0$ and $dR/dn$ are defined in
    Eq.\eqref{eq:fit_staircase}. The error bars correspond to our
    numerics with $1.6 \cdot 10^7$ events.}
  \label{fig:staircase1}
\end{figure}

We observe that photon plus jets events fulfilling
Eq.\eqref{eq:staircuts} alone do not show any kind of simple jet
scaling behavior.  What we are still missing is the crucial photon-jet
separation criterion.  In Fig.~\ref{fig:staircase1} we show two sets
of $\nj$ distributions for different separation criteria. Inspired by
the $W/Z$+jets analysis we can define a wide photon-jet separation in
terms of the invariant mass. In Fig.~\ref{fig:staircase1} we find that
almost prefect staircase scaling appears for minimal values of
$m_{\gamma j} \gtrsim m_Z$, with a very slight degradation for
alternative mass scales. The corresponding description in terms of a
geometric separation leads to very similar results, but only once we
require $R_{\gamma,j}>1$.  This value we can understand from the
typical $m_{\gamma j}$ values combined with $p_T>50$~GeV.
In both cases the first ratio $R_{2/1}$ is notorious, an effect that
has been observed in other channels before~\cite{autofocus}. The
origin lies in a strong PDF suppression of the two-jet
configurations. The price to pay for producing an additional jet,
passing in particular the $p_{T,j}$ criterion, is highest for the
transition $\sigma_{\gamma j} \to \sigma_{\gamma jj}$. The measure is
the relative increase in partonic center-of-mass energy in order to
produce the final state with one additional jet. This relative
increase and the corresponding PDF suppression factor, however,
becomes rather insignificant for higher jet multiplicities.  \bigskip

Setting aside the first entry we can test the quality of staircase
scaling by fitting the form
\begin{equation}
  R_{(n+1)/n} 
  = \frac{\sigma_{n+1}}{\sigma_n}
  = R_0 + \frac{dR}{dn} \; n \; ,
  \label{eq:fit_staircase}
\end{equation}
and determine the slope to compare it to the perfect staircase scaling
prediction $dR/dn=0$. For all curves shown in
Fig.~\ref{fig:staircase1} we find $dR/dn$ in the $0.01 - 0.001$ range,
essentially compatible with zero. The constant values $R_0$ range
around $0.14$, but with a small spread.  In Sec.~\ref{sec:relate} we
will contrast these values with the $W/Z$+jets
cases~\cite{staircase1,staircase2,autofocus}.  While the definition of
the hard process does play a role in determining $R_0$, in
Sec.~\ref{sec:poisson} we will see that the far dominant factor is
$p_{T,j}^\text{min}$ fixed in Eq.\eqref{eq:staircuts}.  \bigskip

\begin{figure}[t]
  \includegraphics[width=0.46\textwidth]{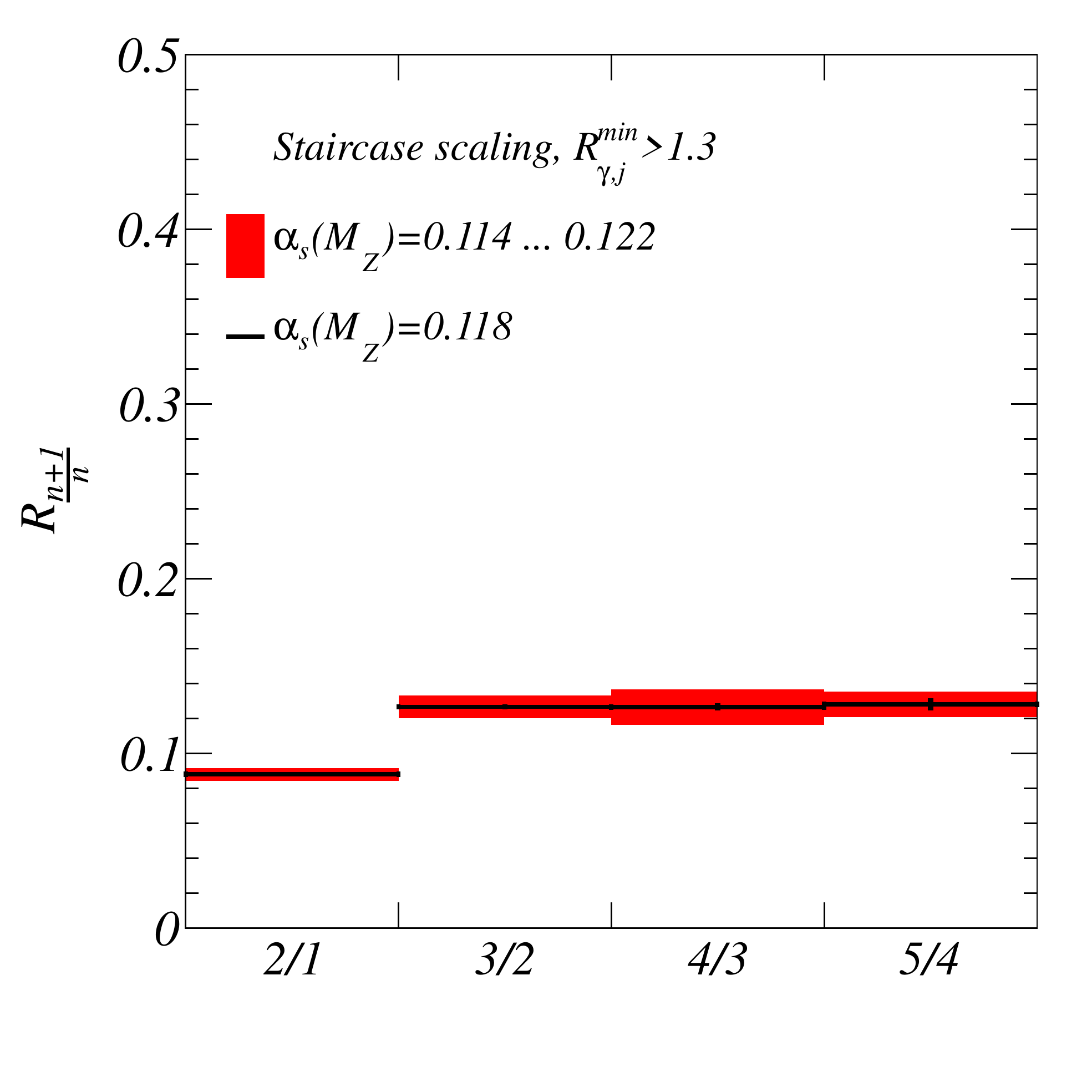}
  \hspace*{0.05\textwidth}
  \includegraphics[width=0.46\textwidth]{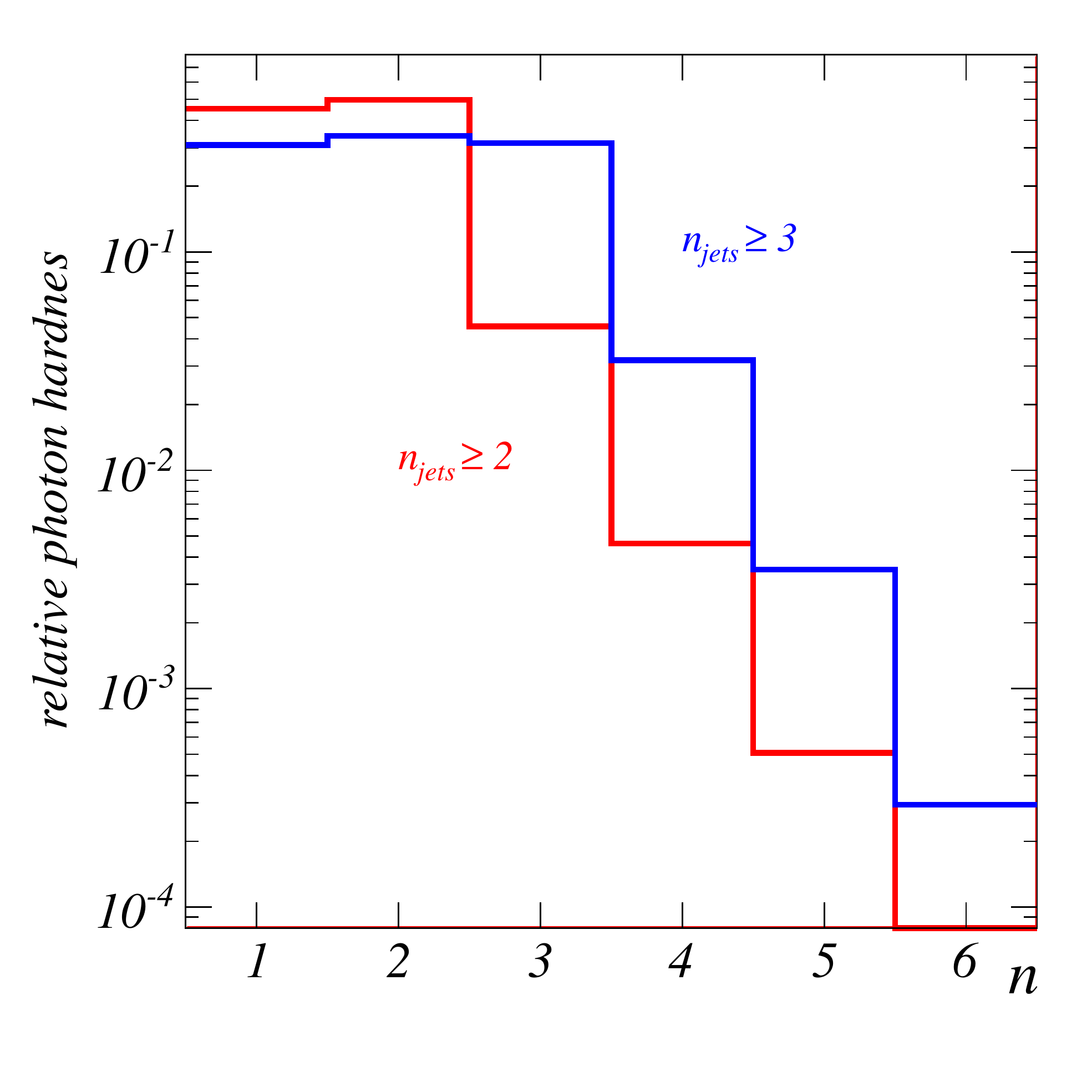}
  \caption{Left: effect of a consistent variation of $\alpha_s(M_{z})$
    on staircase scaling. Right: position of the photon in comparison
    to all jets, ordered by $p_T$. The two curves are for at least
    two-jet and at least three-jet events.}
  \label{fig:staircase2}
\end{figure}

In the left panel of Fig.~\ref{fig:staircase2} we show the effect of
varying $\alpha_s$ on the observed staircase scaling pattern. Between
the largest and lowest values of $\alpha_s$ there is a 7\%
difference. The effect of this shift on $R$ is correspondingly
small. The reason is that $\alpha_s$ and for example the gluon parton
densities are not independently extracted~\cite{cteq10}. An increase
in the value for the strong coupling is compensated by a decrease in
the corresponding gluon density, postponing the expected blow up of
the theory uncertainty to larger $n$ values than we can show in
Fig.~\ref{fig:staircase2}.\bigskip

On theoretical grounds, we can link staircase scaling to the presence
of the large gluon self coupling, \ie the non-abelian structure of
QCD. In the absence of any hard scale from our process, relatively
hard jets are still mostly generated through initial state radiation
(ISR). Our simulation confirms that the final state radiation (FSR)
cascades initiated by the core process jet and additional ISR jets
generates the large and democratic jet multiplicities defining
staircase scaling. This is illustrated in the right panel of
Fig.~\ref{fig:staircase2}. The production of one hard jet with one
hard photon is dominated by the partonic subprocess $qg \to q\gamma$.
Unless we induce a hard scale by cuts, the evolution of the incoming
quark or gluon is dominated by few and non-abelian splittings after
the first hard ISR emission.  Geometrically, for 2-jet events the
quark-photon system typically recoils against the harder of the QCD
jets.  Only from the 3-jet configuration on we can simply split the
ISR gluon, with a jet separation given by $R>0.4$ according to the jet
algorithm. Correspondingly, in the right panel of
Fig.~\ref{fig:staircase2} we see that the photon is typically as hard
as the hardest jets: if we require at least two jets the photon is the
hardest or second hardest object in roughly half of the events each
and the third hardest object only in $\ord(5\%)$ of all events.  For
at least three jets in the final state the photon is equally likely to
be the first, second and third hardest object. In other words, while
one of the jets might usually recoil against a relatively hard photon,
the additional jets responsible for the staircase scaling pattern are
relatively soft. This is a result of splitting hard ISR
gluons. \bigskip

\section{Poisson scaling}
\label{sec:poisson}

According to field theory text book knowledge radiating massless gauge
bosons off, e.g., a hard fermion does not follow a staircase
pattern~\cite{peskin}. Instead, successive soft radiation ordered in
$p_T$ yields a Poisson distribution which can potentially be observed
in the exclusive number of jets
\begin{equation}
  \sigma_n  
  = \sigma_0 \; \frac{e^{-\bar{n}} \; \bar{n}^n}{n!} \; ,
  \label{eq:poisson_define}
\end{equation}
where $\bar n$ is the expected number of emissions. As an example for
such a process consider the exponentiable purely abelian contributions
to multi-jet rates in $e^+e^-$ collisions, see
e.g. \cite{CAVENDISH-HEP-91-5,arXiv:1009.5871}. In this section, we
construct a cut scenario, for which the approximation of subsequent
soft emission is sufficiently realized to observe Poisson scaling in
$\gamma$+jets.

For the exclusive scaling ratios Eq.\gl{eq:poisson_define} translates into
\begin{equation}
  R_{(n+1)/n}
  = \frac{\sigma_{n+1}}{\sigma_n} 
  = \frac{\bar{n}}{n+1} \; .
  \label{eq:poisson_ratio}
\end{equation}
The assumptions entering the derivation of Poisson scaling are
twofold: first, there should be one splitting function, for example
the radiation of a photon or a gluon off a fermion. In the soft limit
successive gauge boson radiation is automatically ordered by the
emission angle. This way, we avoid combinatorial factors of the kind
$n!$ from differently ordered emission in the numerator. The crucial
factor $1/n!$ in Eq.\eqref{eq:poisson_define} appears through the
over-counting of the bosonic phase space. Poisson scaling is what one
expects from a statistical point of view when we assign probabilities
to statistically independent splittings. An example for this
statistical treatment are Sudakov factors or collinear splitting
probabilities following the DGLAP equation. The reason why solutions
of the DGLAP equation show a Poisson behavior is that in its
derivation we only take into account successive splittings of incoming
partons on their way from the proton to the hard process. This is
exactly what corresponds to a resummation of collinear logarithms and
the removal of infrared divergences through the definition of scale
dependent parton densities.  The presence of a large kinematical
logarithm makes parton shower simulations the appropriate tool to
reproduce Poisson scaling.  \bigskip
To force multi-jet scaling for example in association with a photon
into such a Poisson regime we can follow this argument of the parton
shower and the Sudakov factors. What we need is a well defined hard
subprocess, \eg the leading 2-particle $\gamma$-$j_1$ system which
induces many successive splittings of the incoming partons.

\begin{figure}[t]
  \includegraphics[width=0.46\textwidth]{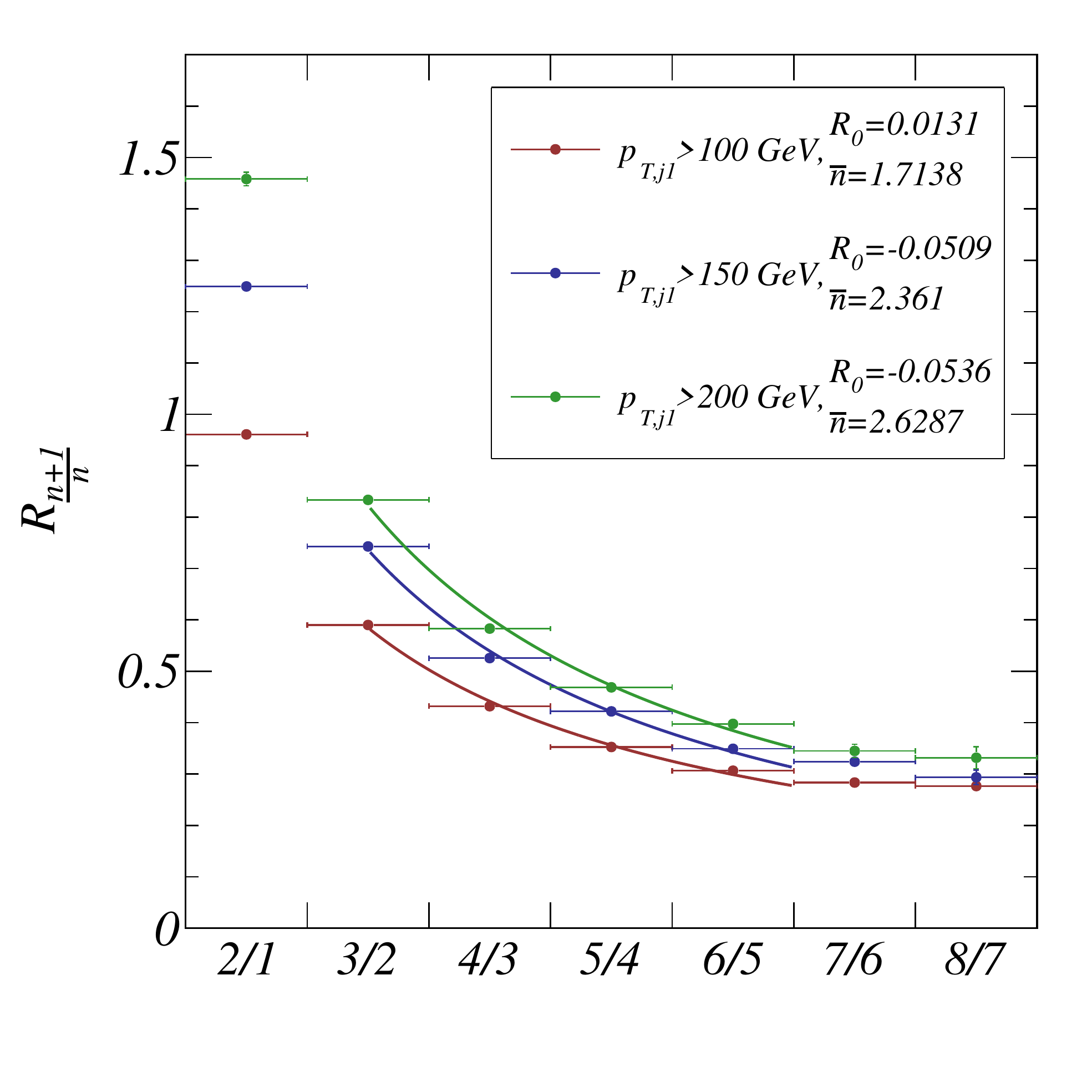}
  \hspace*{0.05\textwidth}
  \includegraphics[width=0.46\textwidth]{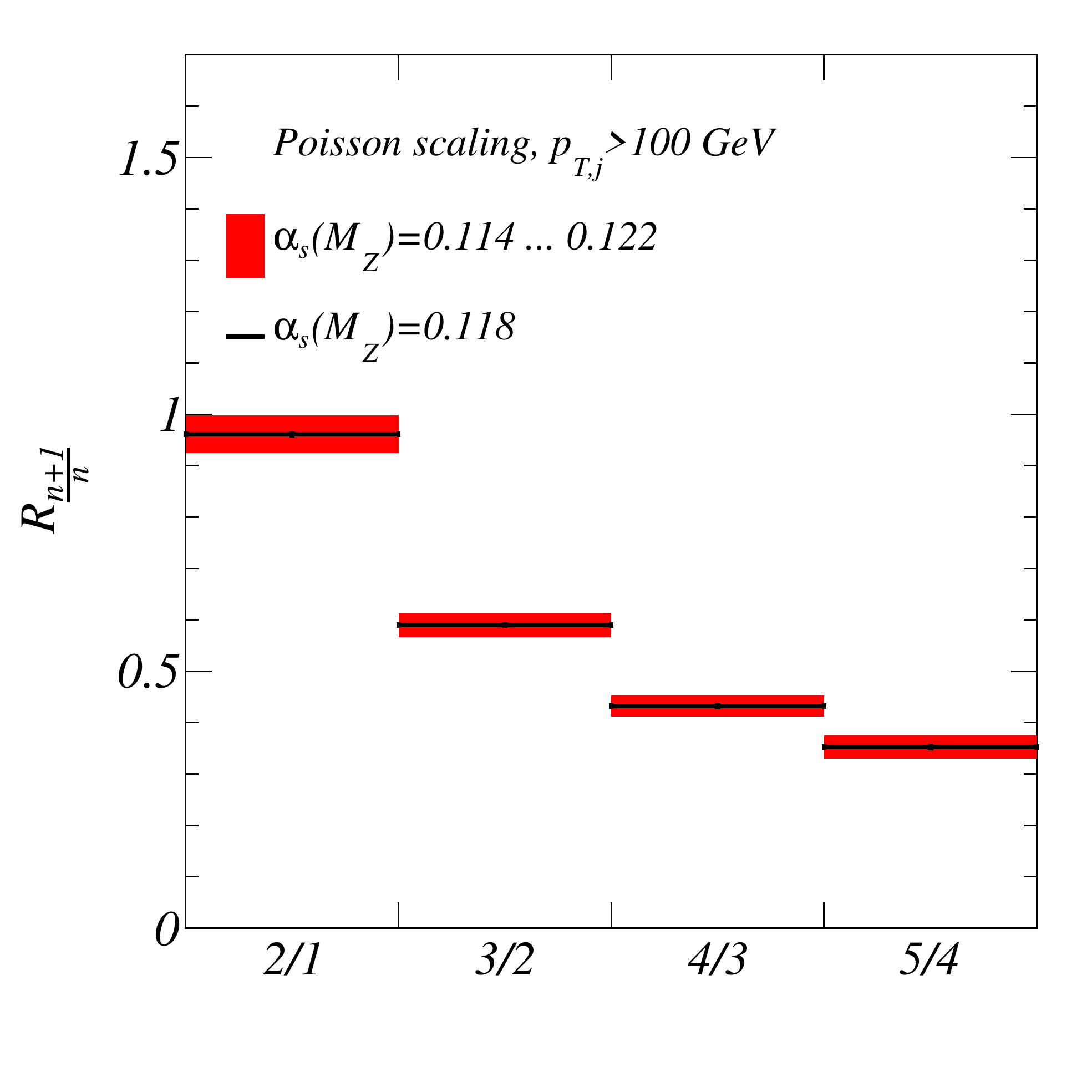}
  \caption{Poisson scaling for $\gamma$+jets production at the 7~TeV
    LHC. Left: different transverse momentum criteria for the leading
    jet, all other jets have $p_{T,j}>20$~GeV. The extracted values
    for $R_0$ and $\bar n$ are defined in Eq.\eqref{eq:fit_poisson}.
    The error bars correspond to our numerics with $2.1 \cdot 10^6$
    events.  Right: effect of a consistent variation of
    $\alpha_s(m_Z)$. }
  \label{fig:poisson1}
\end{figure}

Following the distance measures shown in Fig.~\ref{fig:staircase1} an
obvious choice could be an increased value of $m_{\gamma j_1} \gg
100$~GeV. This is similar to weak boson fusion Higgs production, where
the large invariant mass of the two tagging jets $m_{jj} > 600$~GeV
induces a Poisson scaling of the $Z$+jets
backgrounds~\cite{prl}. However, we find that the cleanest Poisson
distribution is induced by requiring a single hard jet, \ie requiring
\begin{equation}
  p_{T,\gamma} > 20~\gev, 
  \quad 
  p_{T,j} > 100,20,20,...~\gev, 
  \qqqquad 
  |\eta_\gamma| < 2.5, 
  \quad 
  |y_j| < 4.5\,,
  \label{eq:poissoncuts}
\end{equation}
instead of Eq.\eqref{eq:staircuts}. Generating the hard scale through
the hardest jet is more efficient than asking for a hard photon,
because according to our earlier argument a hard photon with
$p_{T,\gamma} = 100-200$~GeV could easily recoil against several jets
from splitting ISR.  Requiring a hard average $p_T$ for the jets would
work as well, though. The important requirement is that through the
cuts we induce a large scale separation with respect to the $p_{T,\rm
  min}$ for radiated additional jets.  \bigskip

In Fig.~\ref{fig:poisson1} we see how enforcing a staggered $p_T$ cut
scenario immediately changes the staircase scaling pattern into a
Poisson distribution. Already for $p_{T,j_1} > 100$~GeV we see a clear
deviation from any kind of staircase behavior provided we allow all
other jets to be as soft as $p_{T,j}>20$~GeV. For $p_{T,j_1} >
150$~GeV the leading $R_{2/1}$ ratio increases to values above unity,
which means that in the exclusive $\nj$ distribution the maximum will
move away from zero.

To test the quality of the Poisson description we fit the
$R_{(n+1)/n}$ distribution which is expected to follow
Eq.\eqref{eq:poisson_ratio}. If we allow for a deviation from the
one-parameter Poisson shape of the kind
\begin{equation}
  R_{(n+1)/n} 
  = \frac{\bar n}{n+1} + R_0 \; ,
  \label{eq:fit_poisson}
\end{equation}
$R_0$ is reminiscent of the staircase pattern
Eq.\eqref{eq:staircase_excl} and should come out essentially zero
while $\bar n$ is the only free parameter in the Poisson shape and
gives the expected number of jets. As expected, the value of $\bar n$
increases for harder leading jets. Just as for the staircase scaling
we do not include the first entry $R_{2/1}$ in the fit.

While it follows the basic expectation, namely becoming large and
exceeding unity, $R_{2/1}$ does not fit the Poisson shape well. In the
first bin we encounter again a PDF suppression effect for producing
the second jet. While this effect is much smaller than the mismatch of
$R_{2/1}$ for staircase scaling, it is important to not include this
bin into the fit to Eq.\eqref{eq:fit_poisson} because it would lead to
obviously wrong best-fit values for $\bar n$ and $R_0$.  \bigskip

If we also exclude the high-multiplicity bins we find small values of
$|R_0|$, closer to zero than to $R_{7/6} \sim R_{8/7}$. This is
expected. Our argument for Poisson scaling rests on the impact of a
large scaling logarithm which has to be generated by successive and
ordered ISR. Beyond some point these successive splittings will stop
feeling a large logarithm and we can expect to fall back onto a
non-negligible staircase scaling.

For example in the case of Higgs production we know that the large-$n$
limits of $R_{(n+1)/n}$ in the staircase and Poisson setups show
hardly any difference. In Fig.~\ref{fig:poisson1} we estimate the
staircase tail to be in the $R_0 \sim 0.3$ range. This is
significantly different from the $R_0 \sim 0.15$ which we find in
Fig.~\ref{fig:staircase1}. The reason is simply the reduced $p_T$
threshold of 20~GeV for the Poisson studies. In the right panel of
Fig.~\ref{fig:poisson1} we again show the consistent variation of
$\alpha_s$. Clearly, the dependence is very small and does not affect
the Poisson scaling feature.

\section{Massive gauge bosons}
\label{sec:relate}

Originally, jet scaling studies have been established for $W$+jets
events. Not surprisingly, $Z$+jets events behave qualitatively and
quantitatively the same~\cite{autofocus}. The purpose of the first two
sections of this paper is to show that after strict jet-photon
isolation we can see the same scaling patterns in photon
production.\bigskip

\begin{figure}[t]
  \includegraphics[width=0.46\textwidth]{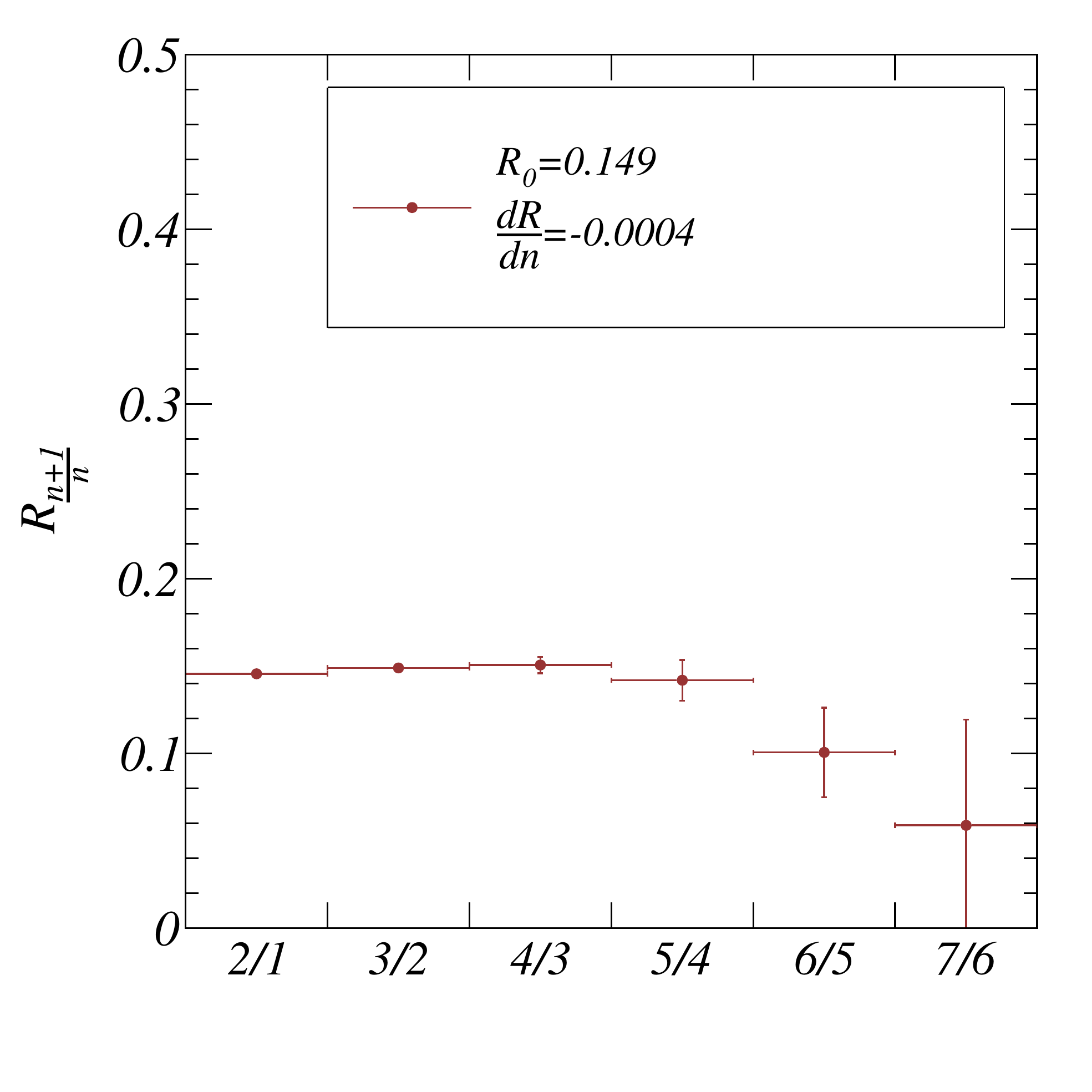}
  \hspace*{0.05\textwidth}
  \includegraphics[width=0.46\textwidth]{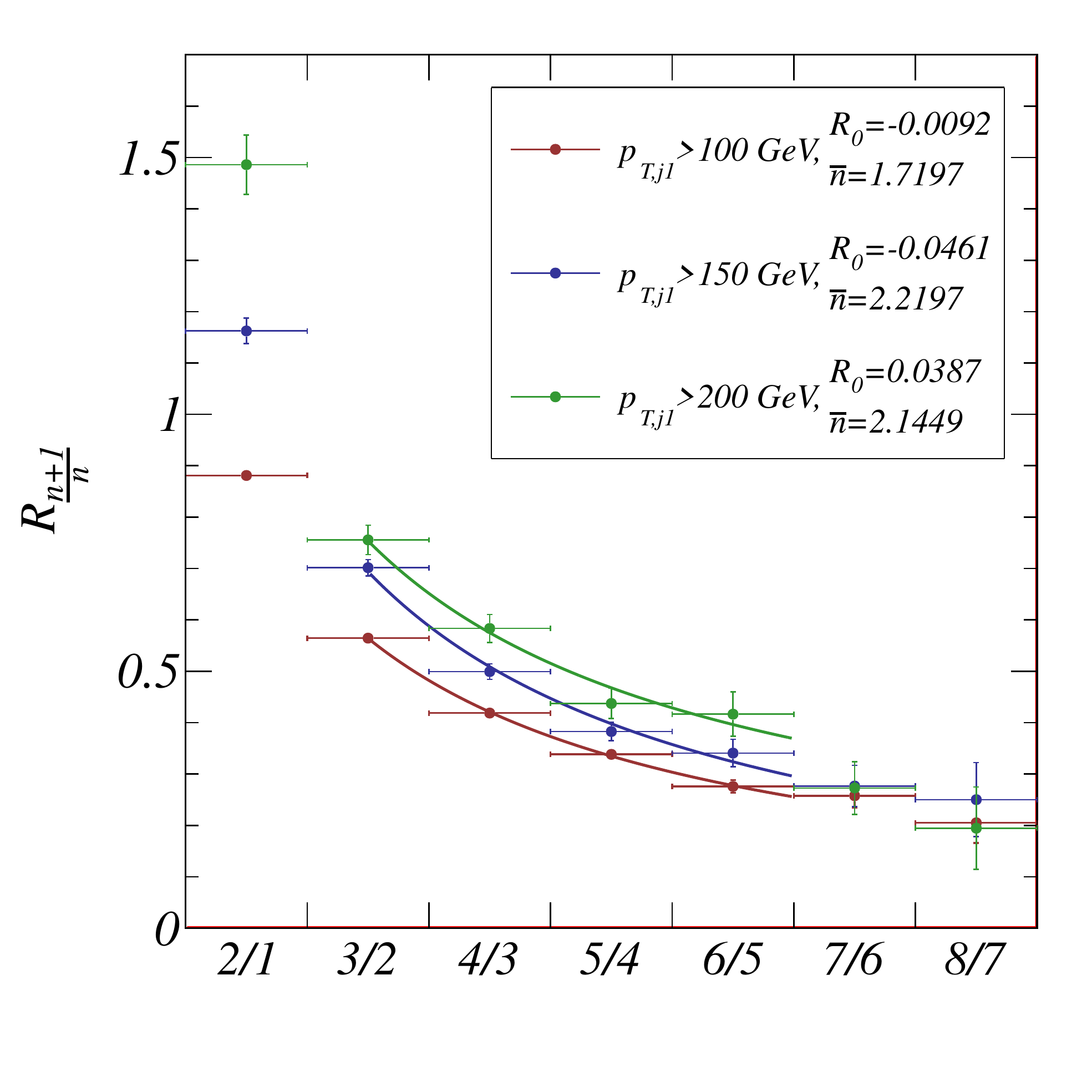}
  \caption{Scaling patterns for $Z$+jets production at the 7~TeV
    LHC. Left: staircase scaling for $p_{T,j}>50~\gev$. Right: Poisson
    scaling for different transverse momenta of the hardest jet, all
    other jets have $p_{T,j}>20$~GeV.  The error bars correspond to
    our numerics with $10^7$ events before cuts.}
  \label{fig:heavy}
\end{figure}

Provided that for well separated photons the non-existence of large
logarithms between inherent mass scales leads to staircase scaling we
expect $\gamma$+jets and $Z$+jets production to be very similar.
This very fact is also exploited experimentally to model the invisibly
decaying $Z$ background to new physics searches \cite{cms} from
corresponding measurements of $\gamma$+jets in control regions
\cite{Bern:2011pa}.
In Fig.~\ref{fig:heavy} we first see that based on the same jet cuts
as in Eq.\eqref{eq:staircuts} $Z$ production follows the same
staircase pattern. The extracted value $R_0 = 0.149$ for
$p_{T,j}>50$~GeV is consistent with the literature~\cite{autofocus} as
well with our findings in Sec.~\ref{sec:staircase}.

For the Poisson regime the situation is slightly different. For the
photons a cut on the leading jet of $p_{T,j_1} > 100$~GeV compared to
a reduced general jet threshold of 20~GeV already induces a large
enough logarithm. The $Z$ mass in the final state could be expected to
further enhance this scaling logarithm, even though it does not really
translate into a collinear logarithm expressed in terms of $p_{T,j}$.
In the right panel of Fig.~\ref{fig:heavy} we find that the $\bar{n}$
values we extract from the $Z$ case are very close to those for the
photons, in particular taking into account the statistical
uncertainties which affect the result for $p_{T,j_1} > 200$~GeV.  The
high-multiplicity staircase limit of the Poisson distribution for $R_0
\sim 0.20-0.25$ again is similar to the photon case.

The agreement of both figures provides a consistency check of our
previous statements but also opens up the possibility to
experimentally cross check $Z$+jets (and $W$+jets) production against
$\gamma$+jets in yet another way.
This holds not only on the qualitative but also on the quantitative
level. The only difference between the two channels is that for
photons the scaling patterns only appear once the photon is very well
separated, controlling any additional QED logarithms which for the
massive $Z$ case do not play any role at these energies.

\section{Outlook}

Counting numbers of exclusive jets has many applications in LHC
searches, implemented for example as distinct Higgs analyses for
different $\nj$ values~\cite{higgs} or central jet
vetos~\cite{wbf}. To apply such cuts while maintaining stability of
the theoretical predictions we need to properly understand exclusive
$\nj$ distributions both experimentally and theoretically.  A major
obstacle to overcome is that exclusive jet measurements have no
straightforward interpretation in fixed order perturbation theory and
exclusive quantities are typically plagued with large theoretical
uncertainties. On the other hand, multi-jet merging, for example using
the {\sc Ckkw} scheme~\cite{ckkw}, allows us to study $\nj$
distributions including a free choice of kinematic cuts and has proven
successful in various experimental analyses so far.\bigskip

Unlike $W/Z$+jets and pure QCD jets production, the associated
production of jets with a hard photon naively does not show simple
scaling patterns. We show that once we require a widely separated
photon we recover staircase scaling $\sigma_{n+1}/\sigma_n = R_0$ for
the total cross section~\hbox{\cite{staircase1,staircase2,autofocus}}.

Once we induce a large logarithm through kinematic cuts we see how the
scaling pattern turns into a Poisson distribution for the exclusive
number of jets. This is known for weak-boson-fusion cuts in Higgs
production~\cite{prl}. For our photon channel a transverse momentum
cut on the leading jet (and not on the photon) works best. Only in the
high-multiplicity regime an underlying 
staircase pattern remains.\bigskip

Given our observations and the large available photon sample at LHC,
this channel is especially well suited to study jet scaling, including a proper experimental and
theoretical error analysis. Combined with more channels where we expect 
staircase scaling, such a comprehensive study will not only provide
crucial information on MC validation but will also help to eventually
reveal the origin of staircase scaling.

A translation of these photon measurements into $W/Z$+jets production
is straightforward and will significantly impact new physics searches:
typically, $\gamma$+jets production is used to infer
$(Z\to\nu\bar\nu)$+jets production in both control and signal
regions~\cite{Bern:2011pa}. Therefore, a better understanding of
exclusive jet quantities and the translation of $\gamma$+jets into
$(Z\to\nu\bar\nu)$+jets production can help to systematically reduce
the uncertainty of the background extrapolation in \eg the
jets+missing energy channel.

\acknowledgments

We are grateful to Alex Tapper for pointing us to this channel.  The
simulations underlying this study have been performed in parts on
bwGRiD, member of the German D-Grid initiative, funded by the
Bundesministerium f\"ur Bildung und Forschung and the Ministerium
f\"ur Wissenschaft, Forschung und Kunst Baden-W\"urttemberg.


\end{document}